\documentstyle[preprint,prb,aps]{revtex}

\begin{document} \draft

\title{Resonant Raman Scattering by Quadrupolar Vibrations of Ni-Ag Core-shell Nanoparticles}

\author{Herv\'e Portales, Lucien Saviot\footnote{Permanent address: Laboratoire de Recherche sur la R\'eactivit\'e des Solides, Universit\'e de Bourgogne et CNRS, BP 47870, 21078 Dijon Cedex, France}, Eug\`ene Duval}
\address{Laboratoire de Physico-Chimie des Mat\'eriaux Luminescents, Universit\'e Lyon I et CNRS, 69622 Villeurbanne Cedex, France}
\author{M\'elanie Gaudry, Emmanuel Cottancin, Michel Pellarin, Jean Lerm\'e, Michel Broyer}
\address{Laboratoire de Spectrom\'etrie Ionique et Mol\'eculaire, Universit\'e Lyon I et CNRS, 69622 Villeurbanne Cedex, France}

\date{\today}

\maketitle

\begin{abstract}

Low-frequency Raman scattering experiments have been performed on thin films consisting of nickel-silver composite nanoparticles embedded in alumina matrix. It is observed that the Raman scattering by the quadrupolar modes, strongly enhanced when the light excitation is resonant with the surface dipolar excitation, is mainly governed by the silver electron  contribution to the plasmon excitation. The Raman results are in agreement with a core-shell structure of the nanoparticles, the silver shell being loosely bonded to the nickel core.

\end{abstract}


\section{INTRODUCTION}

The quadrupolar vibrational modes of silver or gold nanoparticles embedded in dielectric matrices are easily observable by low-frequency Raman scattering (LFRS) because of the enhancement effect when the light excitation is resonant with the dipolar surface plasmon excitation \cite{Mar88,Pal99}. This enhancement will be referred to as "resonant Raman scattering" throughout the paper. For silver particles the absorption by the dipolar plasmon excitation is exhausted by the conduction electrons, and is well separated from the absorption corresponding to the interband transitions. Conversely, for gold nanoparticles the plasmon excitation, in the $450-650\,nm$ spectral range, overlaps the low-energy wing of the interband transitions \cite{Pal98}, so that there is a mixing of both excitations. However, the Raman scattering selects the conduction-electron part of the dipolar plasmon for the resonance enhancement \cite{Por01}. Such a resonance effect was confirmed in the Ag-Au alloy particles in which the contribution of the conduction electrons in the plasmon band increases with the concentration of silver \cite{Por01}.

The study of the Ni-Ag nanoparticles is interesting for two reasons. One the one hand, the dipolar plasmon excitation does not exist in nickel nanoparticles due to the strong extinction coefficient originating from both the intraband and interband transitions. Nevertheless, when adding some silver to the nickel, the contribution of the conduction electrons is expected to lead to a surface plasmon resonance. On the other hand, nickel and silver do not form an alloy, at least in macroscopically sized samples. In consequence, it will be interesting to determine the confinement effect on the spatial distribution of the Ag and Ni atoms and to observe the evolution of the dipolar plasmon excitation as a function of the silver concentration in the nanoparticles. The different points that were addressed in our research are the following: (1) The resonance effect in the Raman scattering from Ni-Ag nanoparticles embedded in dielectric matrix, (2) the spatial distribution of Ni and Ag atoms in the nanoparticles, (3) the vibrational modes in the composite nanoparticles.

\section{EXPERIMENTAL}

\subsection{Preparation and characterization of Samples}

In our experiment, clusters are produced in a laser vaporization source, described in previous papers \cite{Pal98,Pel93}. The beam of a frequency-doubled Nd:YAG pulsed laser ($\lambda = 532\,nm$) is focused onto a bimetallic rod of given $Ni_{x}Ag_{1-x}$ composition ($x = 0.25$ to 1 in step of 0.25). A continuous helium flow cools the laser-produced plasma, and the mixture finally combines into clusters and undergoes a supersonic expansion through a nozzle, at the exit of the source. Moreover, the size distribution and therefore the mean size of the nanoparticles in the sample is controlled by varying the helium pressure. Then, the neutral clusters pass through a skimmer into a vacuum chamber (under approximately $10^{-7}\,mbar$) where they are deposited on a substrate, simultaneously with the alumina matrix. The alumina is evaporated with an electron gun, and the choice of the substrate depends on the characterization experiments to be performed. The metal volumic fraction is taken below $5\,\%$ to minimize the cluster coalescence. The alumina is amorphous and slightly over-stoichiometric in oxygen $(Al_{2}O_{3.2})$, and very porous (about $50\,\%$).

Energy Dispersive X-ray (EDX) and Rutherford Back Scattering (RBS) analyses show that the average stoichiometry of the clusters in the sample is the same as in the alloy rod. Low-energy ion scattering (LEIS) measurements on $Ni_{x}Ag_{1-x}$ clusters show that the silver atoms move onto the surface of the mixed particles \cite{Rou00}. This phenomenon can be explained by the lower surface energy of silver compared to nickel. One can also emphasize that silver and nickel are known to be immiscible metals in the bulk. Transmission Electron Microscopy (TEM) shows almost spherical, and randomly distributed nanoparticles, and allows to determine the size distribution (Fig. 1).

The optical transmission of the films deposited on pure silica (suprasil) has been measured with a double beam Perkin-Elmer spectrophotometer, in the photon energy range $1.45\,eV$ - $4.51\,eV$ ($275\,nm$ - $855\,nm$). Due to the rather low metal volumic fraction in the films the recorded transmission reflects the -- size distribution-averaged -- absorption of the individual particles.

In Fig. 2 are displayed absorption spectra of embedded $(Ni_{x}Ag_{1-x})$ clusters for three different compositions ($x = 0.25$, 0.5, 0.75) and of pure Ni clusters ($x = 1$). This last one does not show any resonance (simple Mie calculations indicate that this holds true for the whole spectral range). When some silver is added to the nickel, a broad band appears on the absorption spectra. This absorption band, which is due to the surface plasmon resonance, becomes very pronounced for silver concentration exceeding $50\,\%$ ($x \leq 0.5$) and the larger the silver concentration in the nanoparticles, the narrower the band plasmon. Moreover, in all the spectra where a plasmon band occurs, one can see that this latter is almost in the same spectral range as the one obtained for pure silver clusters. More details about this study have been given previously \cite{Isspic10}.

\subsection{Raman technique}

The Raman spectra were recorded on a DILOR Z40 monochromator. The high rejection rate and resolution of this setup, which is due to its five gratings, makes it possible to measure a low-frequency signal close to the Rayleigh line. All the visible lines of an $Ar^{+}$ and a $Kr^{+}$ laser were used for excitation. It was controlled that the laser beam does not heat the sample. The incoming light beam is at grazing incidence and the scattered one was detected at $\pi/2$ with respect to excitation.

\section{RAMAN EXPERIMENTAL RESULTS}

No low-frequency Raman scattering was observed from pure Ni nanocrystals in alumina film with blue, green or red excitations. Relatively intense low-frequency peaks appear from Ni-Ag nanoparticles, even with only $25\,\%$ of silver (Fig. 3). As for pure silver nanoparticles \cite{Pal99}, both an intensity decrease and a shift to lower frequencies were observed when the excitation is turned from the blue to the red. These observations clearly illustrate the effect of resonance with the silver conduction-electron contribution to the dipolar plasmon. The one-electron intraband or interband absorption transitions of nickel do not contribute to the resonance enhancement of the Raman scattering by nanoparticle vibrational modes. It was also the case for the interband transitions of gold \cite{Por01}.

The depolarization ratio, $I_{VH}/I_{VV}$, of the Raman intensity for perpendicular polarizations of excitation and scattered lights ($VH$) over the one for parallel polarizations ($VV$) has been measured. $I_{VH}/I_{VV}\simeq 0.65$ by excitation at $514.5\,nm$, and $I_{VH}/I_{VV} \simeq 0.4$ by excitation at $647.1\,nm$. It is well known from group theory and due to the plasmon resonance effect that only the quadrupolar modes of nanoparticles exhibit high Raman activity \cite{Pal99}. For Raman scattering by the five-fold degenerate quadrupolar modes of spherical particles the theoretical depolarization ratio is found to be equal to 0.75 \cite{Mar88,Pal99}. On the other hand,  for resonant Raman scattering by the non-degenerate mode of randomly oriented ellipsoidal particles it is equal to 0.33. By comparison with these theoretical values,  the experimental depolarization ratios show that the nanoparticles which are excited in the green are mainly the spherical shaped ones while they are ellipsoidal in the red, as it was established for pure silver nanocrystals \cite{Pal99}. The Raman scattering by the quadrupolar modes is relatively intense because of the strong coupling between the vibrational quadrupolar modes and the electronic dipolar plasmon in silver nanoparticles \cite{Pal99}. Recently, the Raman scattering by spherical modes (i.e. the breathing mode and its two first harmonics) was observed for silver nanocrystals with narrow size dispersion \cite{Port01}. In the case of nanoparticles with a relatively broad size dispersion, as in this experiment, the Raman peaks corresponding to the scattering by the spherical modes are expected to be of very weak intensity. So, these peaks can be easily masked by the high-frequency tail of the much more intense Raman peak due to the quadrupolar mode, and it is therefore not surprising that they do not appear on our LFRS spectra (Fig. 3).

One observes in Fig. 3 that the position of the Raman peak does not depend very much on the composition of the nanoparticles. This is a priori not consistent with a homogeneous mixing of Ni and Ag atoms, the elastic constants of nickel being approximately two times larger than the silver ones. This is in agreement with the Ni-Ag demixion observed by LEIS \cite{Rou00}. In the following part, the experimental vibrational frequencies will be compared to the calculated ones for demixed nanoparticles consisting of a nickel core and a silver shell. It is expected that the vibrational dynamics of the Ni-Ag core-shell nanoparticles depends crucially on the coupling between the silver shell and the nickel core.

\section{INTERPRETATION}

The resolution of the Navier equation, by using the continuous medium approximation, demonstrates \cite{Lam82,Eri75} that the vibrational frequencies $\nu$ of a spherical  nanoparticle are inversely proportional to the diameter $D$ and proportional to the transversal sound velocity $v_{t}$ :

\begin{equation}
\label{one}
\nu=S\frac{v_{t}}{D}
\end{equation}

\noindent
The factor $S$ depends on the mode and on the ratio of the transversal and longitudinal sound velocities. The frequency $\nu$ corresponds to the Raman shift $\omega$, that is generally expressed in $cm^{-1}$: $\omega = \nu/c$, c being the vacuum light speed. Equation-\ref{one} was experimentally verified by the comparison of the small angle neutron scattering with the Raman scattering by nanocrystals embedded in a glassy matrix \cite{Duv86}. In the following, the frequency calculated from Eq.~(\ref{one}), with $D_{max}$ being the diameter at the maximum of the size distribution, will be compared to the frequency at the maximum of the low-frequency Raman band. The suitability of this procedure lies on the assumption that there is no spatial coherence in the nanoparticle, as it was demonstrated in  previous papers for nanoparticles obtained by the same elaboration technique \cite{Pal99,Duv01}. According to the Lamb's theory \cite{Lam82} on the vibrations of an elastic sphere, the $S$ factor is close to 0.85 and 0.84 for the fundamental quadrupolar mode, in the case of silver and nickel respectively. The used transversal sound velocities for nickel and silver are respectively $v_{t}[Ni] = 3000\,m/s$ and $v_{t}[Ag] = 1660\,m/s$ \cite{Kittel,Fuj91}. Using these values, without taking explicitly into account the bonding strength between Ag and Ni atoms and the inhomogeneous distribution of the two metals in the nanoparticles, it is found by taking the concentration-weighted average sound velocity in Eq.~(\ref{one}), that the vibration frequency of a $Ni_{0.75}Ag_{0.25}$ nanoparticle is 1.3 times larger than that of a $Ni_{0.25}Ag_{0.75}$ particle of same size. The experimental ratio is equal to 0.9, and consequently far from the theoretical ratio.

Faced with this discrepancy and in agreement with the LEIS experiment \cite{Rou00}, it is assumed that the nanoparticles have a core-shell geometry with a Ni core and a Ag shell, as suggested above. The possibility that  the immiscible metals separate upon silver and nickel particle formation can be rejected for two reasons: (1) no LEIS from nickel was observed \cite{Rou00}, (2) the absorption bands (Fig. 2) are much broader than the ones obtained from the simple superposition of the absorptions by separate silver and nickel nanoparticles. The vibration frequencies of the core-shell nanoparticle can be calculated by using the Navier equation with specific boundary conditions (See Appendix).

In the first model, it is assumed that there is continuity of both the vibrational displacement and the force at the Ni/Ag interface, and no force at the nanoparticle/matrix interface (due to the large matrix porosity, the contact between the matrix and the particle is very loose so that the hypothesis of free vibrations is justified). For the studied samples, the calculated vibrational frequencies of the quadrupolar mode are more than two times or even more than three times too large in comparison with the experimental ones (cf. Table). It is worth noting that the calculated values are not far from the frequencies obtained from Eq.~(\ref{one}) by taking the concentration-weighted average sound velocity. This discrepancy between the experimental and calculated frequencies proves that the hypothesis of a strong contact between the Ni core and the Ag shell is not correct.

In the second model, the hypothesis of a core-shell structure is kept, but now it is assumed that the Ag shell boundary at the Ag/Ni interface is free (no force, i.e., no effective contact of the Ag shell with the Ni core). As it can be seen in the Table, the frequencies calculated for the Ag shell quadrupolar vibrations are very close to the experimental ones, while they are about four times larger for the Ni core vibrations. This agreement is a strong indication that the vibrations observed by Raman scattering are those of the Ag shell. This is not surprising, since the Raman process is resonant, and that only the conduction-electrons in silver contribute to the effect of resonance. The free internal boundary of the Ag shell is a priori more questionable. However, this approximation can be justified. Silver and nickel do not form an alloy, this means that the bonding between Ag and Ni atoms is weaker than between same atoms. Moreover, the atomic radii of these metals are different ($r_{Ag} = 1.75$ \AA, $r_{Ni} = 1.62$ \AA), resulting in a mismatch between the Ni and Ag lattices. Furthermore, weak Raman lines corresponding to nickel oxyde were observed and it is likely that nickel atoms at the Ag/Ni interface are oxydized. Nickel is much more oxydizable than silver, so that some nickel atoms can have been oxydized during the deposition process, and likely after deposition  by oxygen diffusion through the porous alumina matrix and the thin silver shell. In consequence, it is expected that the Ag/Ni interface is relatively loose. Moreover, the elastic constants of Ni are approximately three times larger than the Ag ones, so that there exists a large difference between the vibrational frequencies of the Ni core and the Ag shell, and consequently, the phase matching is hardly achieved.

Another type of core-shell nanoparticle was considered to simulate the Ag/Ni interface. Between the Ag shell and the Ni core, a thin shell of a soft X material is placed. Its thickness is chosen to be $5\,\%$ of the core radius. This type of structure is symbolized by Core-X-Shell in the Table. Taking a density equal to $5000\,kg/m^{3}$ and a sound velocity equal to $100\,m/s$ in the X material, and assuming the vibrational and force continuity at the inner interfaces, the calculated frequencies for the fundamental quadrupolar mode are again close to the experimental ones (cf. Table). However, in this model the vibrational frequencies are dependent on both the sound velocity and the density of the X material. The free Ag shell approximation, which does not use adjustable parameters, is convincing enough to allow the interpretation of the low-frequency Raman lines as due to the plasmon-resonant scattering by the quadrupolar vibrations of the free silver shell in Ni-Ag core-shell nanoparticles. Low-frequency Raman scattering was observed from nanoparticles of same composition with smaller diameters ($1.5\,nm$ and $1.2\,nm$). The experimental frequencies are approximately two times smaller than the ones calculated with the free shell model. This discrepancy can be explained by the unsuitability of the continuous medium approximation for so small nanoparticles. Very recently, it was shown by simulation that this approximation becomes unvalid for sizes smaller than some limit and that the sound velocity which has to be taken in Eq.~(\ref{one}) can be much smaller than the bulk one \cite{Wit01}. However, such a simulation, that was carried out for amorphous two-dimensional clusters using  Lennard-Jones  potential, does not allow to estimate this limit for our metallic nanocrystals.

\section{CONCLUSION}

The observation of low-frequency Raman scattering by the quadrupolar vibrational mode of nickel-silver composite nanoparticles confirms that there is an effect of resonance with the silver conduction-electron contribution to the surface dipolar plasmon excitation. No low-frequency Raman scattering was observed from pure nickel nanoparticles. The frequencies of the vibrational modes are in agreement with a core-shell structure of the particles. The Raman experiment is quantitatively well interpreted by assuming that the scattering originates from the vibrations of a silver shell loosely bonded to the nickel core.

\appendix

\section{Free vibrational modes of core-shell, shell and onion-like systems}

The calculations presented here are based on Lamb's approach \cite{Lam82} and on Eringen and Suhubi's equations \cite{Eri75}. These authors express in spherical coordinates the displacement vector $\overrightarrow u(\overrightarrow r,t)$ as a function of the position vector $\overrightarrow r$ and time $t$, and the force $\overrightarrow F(\overrightarrow r,t)$ that is applied to the element area perpendicular to  $\overrightarrow r$. In this paper, the following vectors are used ($\overrightarrow{e_r}$, $\overrightarrow{e_\theta}$ and $\overrightarrow{e_\varphi}$ being the usual spherical unit vectors):

\begin{equation}
\left\{
  \begin{array}{rcccccc}
  \overrightarrow{e_1} & = & Y^m_l \left( \theta, \varphi \right) \overrightarrow{e_r}\\
  \overrightarrow{e_2} & = & \frac{\partial Y^m_l \left( \theta, \varphi \right)}{\partial \theta} & \overrightarrow{e_\theta} & + &\frac{1}{\sin \theta} \frac{\partial Y^m_l \left( \theta, \varphi \right)}{\partial \varphi} &\overrightarrow{e_\varphi}\\
  \overrightarrow{e_3} & = & \frac{1}{\sin \theta} \frac{\partial Y^m_l \left( \theta, \varphi \right)}{\partial \varphi} &\overrightarrow{e_\theta} & - &\frac{\partial Y^m_l \left( \theta, \varphi \right)} {\partial \theta} &\overrightarrow{e_\varphi}\\
  \end{array}
\right.
\label{base}
\end{equation}

It should be noted that $\overrightarrow e_2 = \overrightarrow e_3 = \overrightarrow 0$ for $l=0$.

The projections of $\overrightarrow u$ and $\overrightarrow F$ along these three vectors are given in Eqs.~(\ref{u}) and (\ref{F}) respectively.

\begin{equation}
\left\{
  \begin{array}{rcl}
  u_1 & = & \frac1r \left\{ A \left[ l z_l\left( q r \right) - q r z_{l+1} \left( q r \right) \right] + C \left[ l \left( l+1 \right) z_l \left( Q r \right) \right] \right\} \exp \left( -i \omega t \right)\\
  u_2 & = & \frac1r \left\{ A z_l \left( q r \right) + C \left[ \left( l+ 1 \right) z_l \left( Q r \right) - Qr z_{l+1} \left( Q r \right) \right] \right\} \exp \left( -i \omega t \right)\\
  u_3 & = & B z_l \left( Q r \right) \exp \left( -i \omega t \right)\\
  \end{array}
\right.
\label{u}
\end{equation}

\begin{equation}
\left\{
  \begin{array}{rcl}
  F_1 & = & \frac{2 \mu}{r^2} \left\{ A \left[ \left( l^2-l- \frac{Q^2 r^2}{2} \right) z_l \left( q r \right) + 2 q r z_{l+1} \left( q r \right) \right] + C l \left( l+1 \right) \left[ \left( l-1 \right) z_l \left( Q r \right) - Q r z_{l+1} \left( Q r \right) \right] \right\}\\
      &   & \exp \left( -i \omega t \right)\\
  F_2 & = & \frac{2 \mu}{r^2} \left\{ A \left[ \left( l-1 \right) z_l \left( q r \right) - q r z_{l+1} \left( q r \right) \right] + C \left[ \left( l^2-1- \frac{Q^2 r^2}{2} \right) z_l \left( Q r \right) + Q r z_{l+1} \left( Q r \right) \right] \right\} \exp \left( -i \omega t \right)\\
  F_3 & = & B \frac{\mu}{r} \left[ \left( l-1 \right) z_l \left( Q r \right) - Q r z_{l+1} \left( Q r \right) \right] \exp \left( -i \omega t \right)\\
  \end{array}
\right.
\label{F}
\end{equation}

\noindent
where $q = \omega / v_l$, $Q = \omega / v_t$, ($v_l$ and $v_t$ are the longitudinal and transversal sound velocities), and $\mu$ is one of the Lam\'e's constants. $z_l$ is a Bessel function. Because $\overrightarrow u$ and $\overrightarrow F$ cannot diverge, only Bessel functions of the first kind are used in space regions containing the origin ($z_l \equiv j_l$). Therefore, three constants determine the vibration inside these regions. In regions which do not include the origin, the two kinds of Bessel functions are valid; $z_l$ is a linear combination of $j_l$ and $y_l$, and then six constants are involved. It should be noted that the constant $B$, on which depend $u_3$ and $F_3$, is related to torsional modes. Starting from now, these modes will not be discussed ($B \equiv 0$) because they are not Raman active \cite{Duv92}.

Dealing with spheres, shells, core-shells or onions, is done by changing the boundary conditions at the various interfaces. Different kinds of boundary conditions are used. For free surfaces we use $\overrightarrow F = \overrightarrow 0$ and for interfaces between two different materials, $\overrightarrow u$ and $\overrightarrow F$ are continuous. Resolving this problem consists in calculating the unknowns in each region.
\begin{itemize}
  \item $l \neq 0$: there are two unknowns for the region containing the origin and four for other regions. Each boundary results in two linear equations if a free surface is involved and four linear equations otherwise.
  \item $l = 0$: ($\overrightarrow e_2 = \overrightarrow e_3 = \overrightarrow 0$) there is one unknown for the region containing the origin and two for other regions. Each boundary results in one linear equation if a free surface is involved and two linear equations otherwise.
\end{itemize}

It is straightforward to check that in each case the system consists in $N$ linear equations with $N$ unknowns. This system admits non-zero solutions when its determinant vanishes. This determinant is a function of $\omega$ only. Therefore, calculating the vibration eigenfrequencies consists in finding the roots of this function. This was performed numerically \cite{Footnote}.

\begin{figure}
\label{f1}
\caption{Diameter distribution of Ni-Ag nanoparticles: (a) $Ni_{0.75}Ag_{0.25}$, (b) $Ni_{0.50}Ag_{0.50}$, (c) $Ni_{0.25}Ag_{0.75}$. The full lines are the log-normal fits.}
\end{figure}

\begin{figure}
\label{f2}
\caption{Absorption coefficient as a function of the wavelength for metallic nanoparticles in alumina films: pure Ni (thick line), $Ni_{0.75}Ag_{0.25}$ (dotted line), $Ni_{0.50}Ag_{0.50}$ (dashed line) and $Ni_{0.25}Ag_{0.75}$ (thin line).}
\end{figure}

\begin{figure}
\label{f3}
\caption{Low-frequency Raman spectra recorded under 457.9 nm laser excitation. $Ni_{0.75}Ag_{0.25}$: squares; $Ni_{0.50}Ag_{0.50}$: circles; $ Ni_{0.25}Ag_{0.75}$: triangles.}
\end{figure}

\begin{table*}
\caption{Comparison of the Raman peak frequencies ($cm^{-1}$) obtained in the different models with the experimental ones for $Ni_{0.75}Ag_{0.25}$ ($D_{max} = 2.3\,nm$), $Ni_{0.50}Ag_{0.50}$ ($D_{max} = 3\,nm$) and $Ni_{0.25}Ag_{0.75}$ ($D_{max} = 2.8\,nm$).}
\begin{tabular}{cccc}
Model or experiment & $Ni_{0.75}Ag_{0.25}$ & $Ni_{0.50}Ag_{0.50}$ & $Ni_{0.25}Ag_{0.75}$\\
\hline
Core-Shell & 34.5 & 24.1 & 21.4\\
Free Ag shell & 10.0 & 8.7 & 11.2\\
Core-X-Shell & 9.98 & 8.2 & 11.3\\
Experiment & 10.1 & 8.2 & 8.8\\
\end{tabular}
\end{table*}


\begin{references}
\bibitem{Footnote}The C source program used in this work is available upon request.
\end{references}
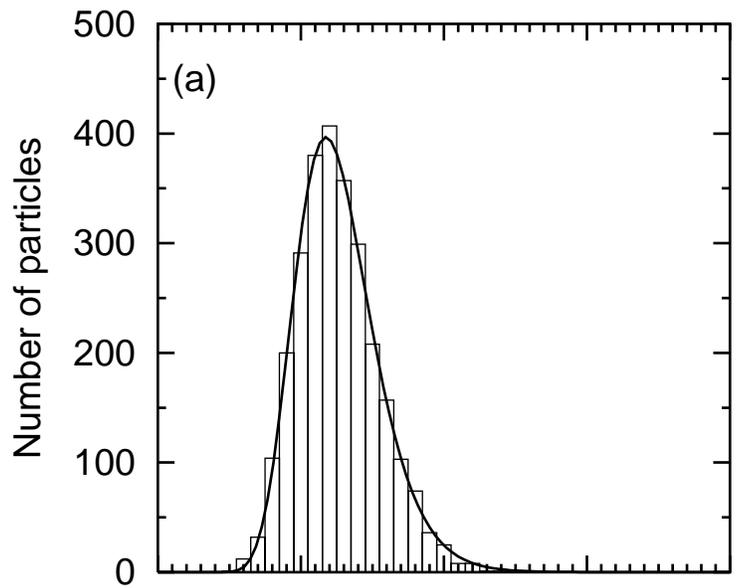
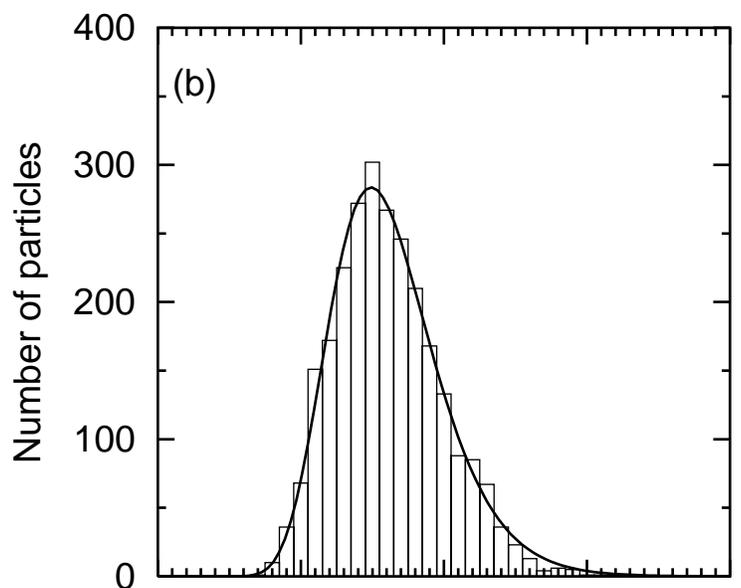
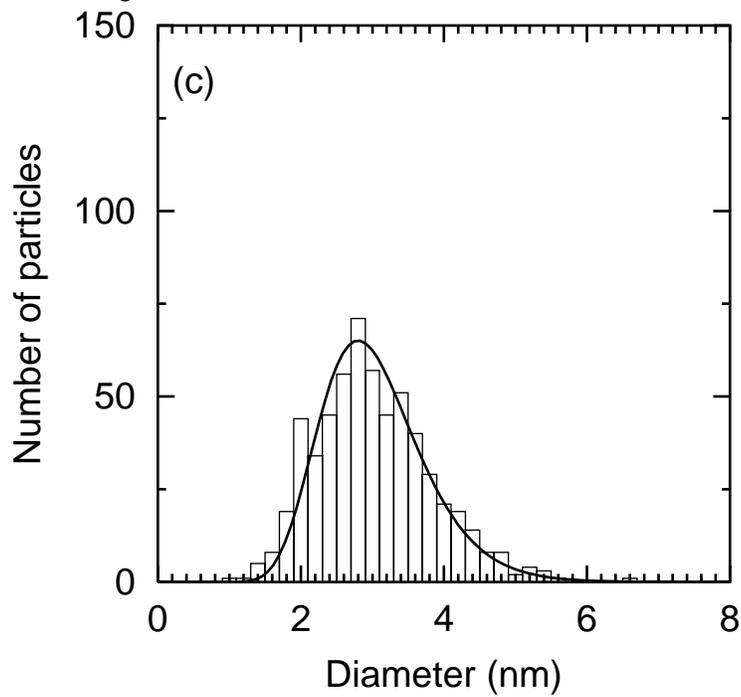
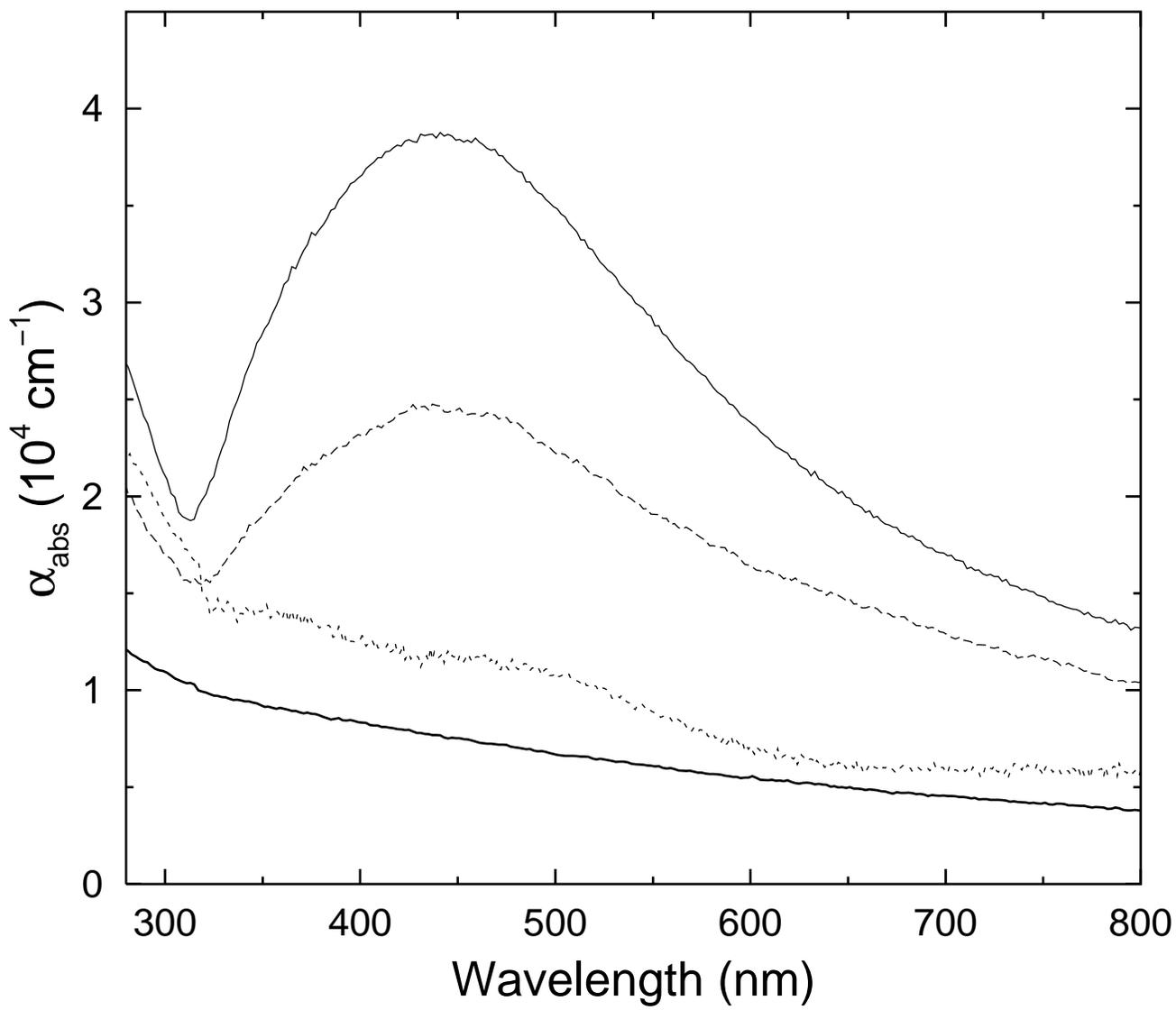
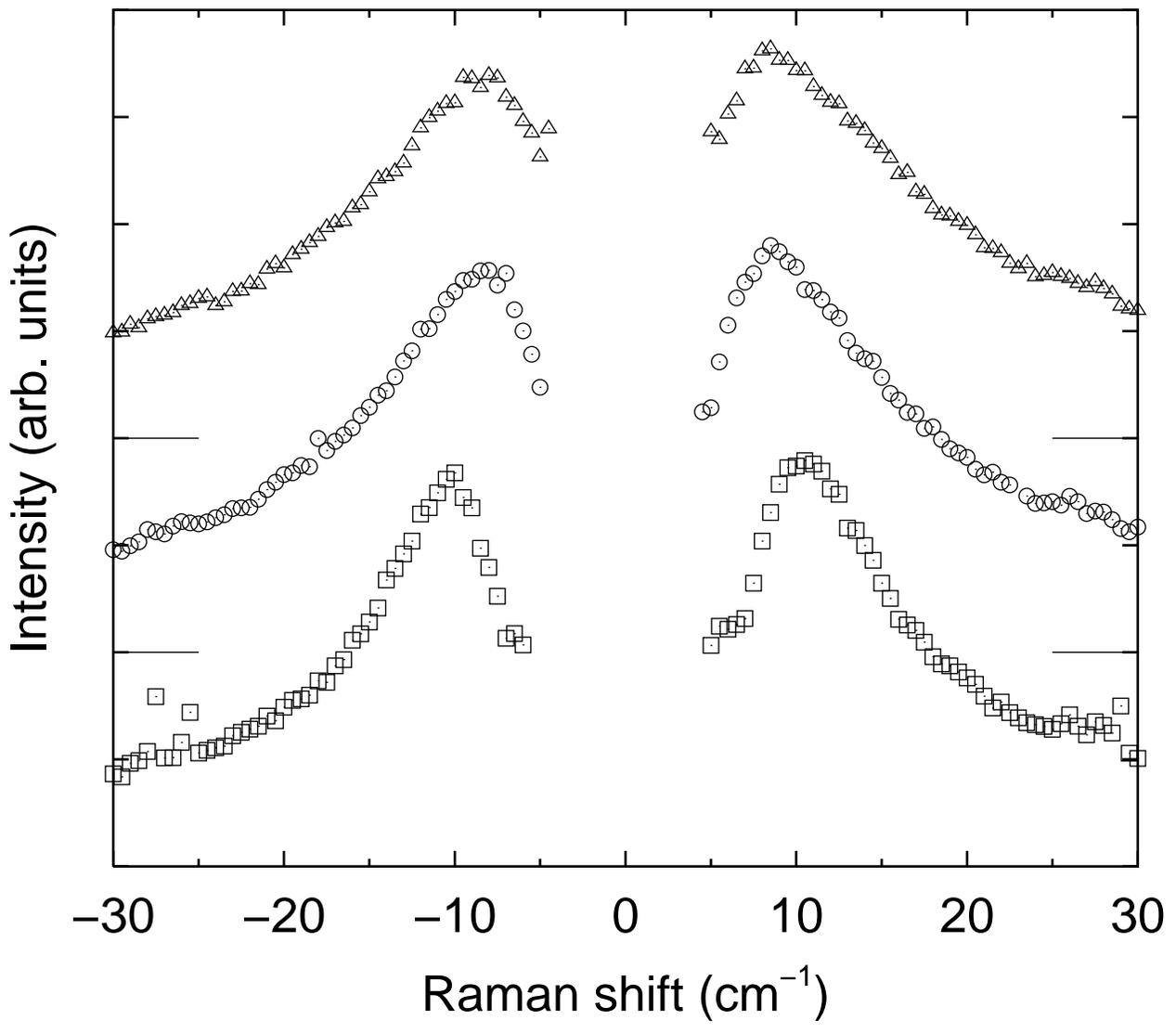
\end{document}